\documentclass{elsarticle}

\usepackage{amssymb}
\usepackage{cool}
\usepackage{float}
\usepackage{caption}
\usepackage{subcaption}
\usepackage{ragged2e}
\usepackage{color} 
\usepackage{a4wide}
\usepackage[hidelinks]{hyperref}

\journal{}

\begin{document}

\begin{frontmatter}



\title{Investigation of Submergence Depth and Wave-Induced Effects on the Performance of a Fully Passive  Energy Harvesting Flapping Foil Operating Beneath the Free Surface.}


\author[inst1]{Nikolaos Petikidis}
\ead{nikos.petik@hotmail.com}
\author[inst1]{George Papadakis\corref{mycorrespondingauthor}}
\ead{papis@fluid.mech.ntua.gr}
\affiliation[inst1]{organization={National Technical University of Athens, School of Naval Architecture \& Marine Engineering},
            addressline={9, Heroon Polytechniou Str}, 
            city={Athens},
            postcode={15780}, 
            country={Greece}}

\cortext[mycorrespondingauthor]{Corresponding author}
\begin{abstract}
This paper investigates the performance of a fully passive flapping foil device for energy harvesting in a free surface flow. The study uses numerical simulations to examine the effects of varying submergence depths and the impact of monochromatic waves on the foil's performance. The results show that the fully passive flapping foil device can achieve high efficiency for submergence depths between 4 and 9 chords, with an "optimum" submergence depth where the flapping foil performance is maximised. The performance was found to be correlated with the resonant frequency of the heaving motion and its proximity to the damped natural frequency. The effects of regular waves on the foil's performance were also investigated, showing that waves with a frequency close to that of the natural frequency of the flapping foil aided energy harvesting. Overall, this study provides insights that could be useful for future design improvements for fully passive flapping foil devices for energy harvesting operating near the free surface.

\end{abstract}




\end{frontmatter}

\graphicspath{{graphs/}} 


\section{Introduction} 

The technology for flapping foils was inspired by fish locomotion. Evolution through hundreds of millions of years has enabled fishes to develop highly efficient thrusting mechanical systems. Except from thrust production, these devices can also be used to capture energy from water currents, such as tidal currents or from rivers.  When the pitching amplitude is higher than a threshold value, for a given frequency and heave amplitude, called the feathering limit, the foil switches from the thrust production mode to the energy harvesting mode. Their main competitors are horizontal axis turbines and vertical axis turbines. Compared to these types of turbines, flapping foils have certain benefits. Firstly, the efficiency of flapping foils remains high even in very large angles of attack with the flow, due to the phenomenon of dynamic stalling which insures that lift remains temporarily high \cite{kiefer2020dynamic}. This also means that oscillating hydrofoils can harvest energy from a wider range of current velocities \cite{young2014review} compared to conventional turbines with a specific design point. The device is also more robust, as centrifugal stresses are absent, and more environmentally friendly, as blade tip velocities are lower (\cite{dai2017optimal} and \cite{domenech2018marine}). Finally, their rectangular sweeping profiles are perfect for shallow and wide fluid channels.

Energy harvesting foils were first proposed by McKinney and DeLaurier in 1981 \cite{mckinney1981wingmill}. It has been shown that flapping foils can offer high performance for both the production of thrust or energy harvest. Many studies have taken place, either experimental (\cite{schouveiler2005performance} and \cite{andersen2017wake}) or numerical (a collection can be found here in  \cite{young2014review}), showing the applicability of these devices and highlighting their advantages and disadvantages over conventional devices. Incorporating chord-wise flexibility also seems promising as increases in performance are significant, as shown in \cite{anevlavi2020non}. Importantly, this device can be simplified by attaching the heaving degree of freedom (DOF) to a spring and damper, and imposing the motion of the pitching DOF, creating the semi-passive flapping foil \cite{zhu2009modeling}. The pitching motion creates a variable lift force that makes the foil oscillate in its heaving DOF. Alternatively, both heaving and pitching DOFs can be attached to springs and dampers, creating the fully passive flapping foil, which oscillates without external forcing (not even initially) when it faces fluid flow. By suitably tuning the structural parameters, a periodical, high energy content in the heaving DOF, motion can be sustained.  \cite{peng2009energy} explored the modes that are achievable by different choices of structural parameters. These fully passive flapping-foils proved to be a viable alternative to the active or semi-passive flapping foils that require intricate mechanisms to enforce the movement. These extra mechanisms raise the risk of failure, the construction cost, and the cost of maintenance. They also have additional power losses. Most research in the past has shown that the fully passive foil exhibits a lower performance compared to the active ones. Recent numerical experiments however, by \cite{boudreau2020parametric}, aiming to optimize the device by exploring the available domain of the structural parameters showed that the fully passive flapping foil can perform as well as active flapping foils while maintaining a much simpler structure. In addition, it was shown that good performance is available for a wide range of parameters, which is necessary for a practical application. 

While a plethora of research has been conducted for active energy harvesting flapping foils under more realistic conditions, not so much has been conducted for fully passive flapping foils. Confinement and 3D effects were studied by \cite{gunther2022confinement}, and the effects of flow perturbation by other flapping foils upstream by \cite{oshkai2022reliability}. However there is a lack of research for the influence of the free surface, which is important if we want a more realistic assessment of the device as flapping foils will likely be placed close to it, for example due to the need to operate in shallow waters, either in rivers or relatively close to shores. The effects of proximity to the free surface, for instance, or the effects of waves, have not yet been explored. The effects of the free surface on active flapping foils were studied by \cite{deng2022effects} and \cite{zhu2022energy}, whose results were similar, finding that the effect of the free surface on the performance of the devices was mostly negative, but becomes quickly irrelevant as submergence depth increases. 
 
With regards to the influence of a wavy free surface on flapping foils, most research has focused so far on thrust producing flapping foils. In  \cite{isshiki1982theory} they was the first to propose that a flapping foil on the bow of a ship could convert wave power into propulsive power and proved this case. The effect of regular waves on thrust producing flapping foils, both rigid and flexible, has been studied by \cite{liu2018effects}, where it was found that power output is significantly increased, especially when the frequency of the waves matches that of the foil and the phase difference is suitable. Efficiency was higher when a flexible foil was used, but was not affected that greatly, in general. An investigation into 3D effects and free surface wave patterns was done by \cite{lopes2023modelling}. In  \cite{xu2017propulsion} they studied thrust producing flapping foils in waves, for constant submergence depth and wave heights using a boundary element method. It was found that an increase in wave frequency, increased the power output but decreased the efficiency. It is distinctive that the amplitudes of the motion fluctuated periodically when the frequencies of the foil and waves did not match. When the frequencies did match there were no fluctuations, as the phase difference of foil and wave remains constant throughout the periods, but the phase difference played a significant role. In \cite{filippas2018semi} they studied a semi-active flapping foil operating in waves and currents, in a shear flow with variable bathymetry (simulating nearshore conditions) using BEM analysis. For a specific moderate wave frequency, significant efficiency can be achieved both when the foil and wave frequency match or don’t, by operating at appropriate pitch angles. Peak efficiency was achieved, however, even for small pitching angles, when the foil matched the frequency of the waves or was double that, and the phase difference was suitable for each case. 

This work is a continuation of the work of \cite{THEODORAKIS2022103674}, who studied how shear present in the flow affects the fully passive flapping foil device, and began the examination of the influence of the free surface. Particularly the influence of a calm free surface is studied in depth, for a broad range of submergence depths and multiple Froude numbers. An initial step in the examination of the impact of monochromatic waves is also taken, examining cases when the wave frequency is close to the flapping foil's.

This paper is structured as follows: In \autoref{sec:num} the in-house Computational Fluid Dynamics solver MaPFlow used for the numerical analysis is outlined briefly. Its strong coupling to a Rigid Body Dynamics solver to simulate the dynamics problem is also detailed. In \autoref{sec:numsetup}, the physical problem is explained and the numerical setup of the solver is given. In \autoref{sec:res} the effect of submergence depth is explored for various Froude numbers. Finally, the passive foil is investigated operating under monochromatic waves of different frequencies and their effect on the foil performance is presented.

\section{Numerical Methodology}\label{sec:num}

The numerical investigation of Fluid Structure Interaction (FSI) problems is performed by combining two separate computational algorithms. Firstly, a flow solver is utilised to describe the fluid motion, and secondly a dynamic solver that computes the structure's response under the flow excitation. The flapping foil is considered submerged and thus the presence of the  free surface requires the solution of a two-phase problem. 
This methodology is implemented as part of the CFD code MaPFlow. The in-house code is developed in NTUA  \cite{Papadakis2014}, \cite{Diakakis2019}, and has proved capable of handling both compressible and purely incompressible flows on arbitrary polyhedral meshes. The code is able to perform in a multi-processing environment utilising the MPI protocol, while the grid partitioning is performed using the Metis Library \cite{Karypis2013}. The numerical methodology is thoroughly described in \cite{THEODORAKIS2022103674}, so here it is briefly reviewed.

\subsection{Governing equations}
 
 The approach used in this work to solve the incompressible flow equations is based on the the artificial compressibility method (ACM) introduced by \cite{Chorin1967}. This approach is  coupled with the Volume of Fluid (VoF) method\cite{Hirt1981}, which tracks the interface between different fluid phases. The ACM solves the unsteady system of equations using the dual--time stepping technique \cite{JAMESON1991}, where a pseudo--steady state problem is solved at each real time iteration. The original unsteady system of equations is augmented with pseudo-time derivatives of the unknown variables, and convergence is achieved when these derivatives approach zero, yielding the original system of equations.

The ACM assumes a relation between the pressure and the density field during pseudo-time, with the blending performed through a numerical parameter $\beta$, where $\frac{\partial\rho}{\partial p}\big|_\tau=\frac{1}{\beta}$. The value of $\beta$ is typically between 5 and 10 for free surface flows \cite{ntouras2020coupled, Dudley2002}.

The governing equations for the two-phase incompressible flows solved using the ACM and VoF methods are described by Equation \eqref{int_govEqs}.
 \begin{equation}
        \Gamma \frac{\partial}{\partial \tau} \int_{D_{i}} \vec{Q}\mathrm{d}D  + 
        \Gamma_{e} \frac{\partial }{\partial t} \int_{D_{i}} \vec{Q}\mathrm{d}D + 
        \int_{\partial D_{i}} \left( \vec{F}_{c} - \vec{F}_{v}\right) \mathrm{d}S 
        = \int_{D_{i}} \vec{S}_{q} \mathrm{d}D
        \label{int_govEqs}
\end{equation}

The above system of equations expresses the change of the primitive variables $\vec{Q}$ inside a control volume $D_i$ with boundary $\partial D_i$, in time $t$. The vector $\vec{Q}=[p,\vec{\upsilon},\vec{\alpha_l}]^T$, includes the pressure $p$, the 3-dimensional velocity vector $\vec{\upsilon}$, and the volume fraction $\alpha_l$. The volume fraction $\alpha_l$ indicates the presence of either the liquid phase with density $\rho_l$, or the presence of the gaseous phase with density $\rho_g$. Using the volume fraction the density of the mixture can be found as $\rho_m = \alpha_l\rho_l+(1-\alpha_l)\rho_g$.\par

For the turbulence closure, the k-$\omega$ SST model of \cite{Menter1994} is employed. In case of free surface flows, it has been noted that the turbulence models tend to overproduce turbulence viscosity in the vicinity of the free surface \cite{Larsen2018}, \cite{Kamath2015}. In order to suppress the turbulent viscosity near the free surface \cite{Devolder2017} introduced a source term in the equation of the turbulent kinetic energy. This source term is activated near the free surface and scales with the local viscosity and the gravity vector.
To simulate numerical wave tanks, damping or wave generation zones are implemented at the lateral boundaries of the computational domain in order to absorb or create free surface waves and prevent reflections due to boundary conditions. In MaPFlow, the forcing zone technique is used to achieve this by adding source terms to the momentum equations. These source terms are added only to the momentum equations and take the form of Equation \eqref{SUnwt}, where $C_{nwt}$ is a coefficient used to smoothly vary the influence of the forcing terms from the start of the forcing zone to its end. The damping is performed by eliminating the vertical component of the velocity vector, while the generation of waves is achieved by forcing the numerical solution according to the specified wave theory.

\begin{equation}
	\vec{S}_{nwt}= C_{nwt}\rho_{m}\left( \vec{\upsilon}-\vec{\upsilon}_{tar} \right)
	\label{SUnwt}
\end{equation}

The coefficient $C_{nwt}$ is regulated by the factor $\alpha_{nwt}$ and a function $f_{nwt}$, which is defined inside the forcing zones. The smooth transition from the start to the end of the forcing zone is controlled by the parameter $x_{r}$, as shown in Equation \eqref{Cnwt}. 
The influence of the various parameters of the coefficient is examined in \cite{ntouras2020coupled, Peric2015}.
\begin{equation}
	C_{nwt}	= \alpha_{nwt} f_{nwt}(x_r)\:, \:\: x_{r}=\frac{x_{s}-x}{x_{s}-x_{e}}
	\label{Cnwt}
\end{equation}

The function $f_{nwt}$ takes the form of an exponential function, as shown in Equation \eqref{Fnwt}.

\begin{equation}
   f_{nwt}(x_r) = \frac{\exp\left( x_{r}^{n} \right)-1}{\exp\left( 1 \right)-1}
   \label{Fnwt}
\end{equation}

\subsection{Discretization}

The finite volume method is employed to discretize the equations on a computational mesh, and the dual-time stepping technique is used to solve the equations implicitly in pseudo-time (denoted by $\tau$). The approximate Riemann solver of Roe \cite{Roe1981} is used to evaluate the convection terms, and the second-order central differentiation scheme is used for the viscous fluxes. To facilitate convergence the system of equations is equipped with the Kunz-preconditioner \cite{Kunz2000} (denoted by $\Gamma$ in \eqref{int_govEqs}). Regarding spatial discretization, the reconstruction of the velocity field uses a standard piecewise linear interpolation scheme \cite{ntouras2020coupled}, while the reconstruction of the pressure is based on \cite{Queutey2007}. For the volume fraction reconstruction, the BICS scheme \cite{Wackers2011} is used.

In this work to account for the foil's motion a deforming grid algorithm is employed. Consequently, to ensure mass conservation, the Geometric Conservation Law (GCL) must be satisfied \cite{Ahn2006}. This is achieved including a residual term in the time discretization of the unsteady equations. Finally, for the  the grid deformation a radial basis functions (RBF)  approach is employed (see \cite{RendallAllen2008}).

Regarding time discretization, two successive levels of solution are retained, yielding a second order accurate scheme while the ~fictitious  is discretized using a first-order  backward difference~scheme \cite{Biedron2005} .

\subsection{Fluid Structure Interaction}
This section briefly describes the fluid-structure interaction (FSI) methodology. The energy harvesting foil is fully passive and thus it's motion is driven by the forces exerted on it by the surrounding fluid. The study focuses on 2D approach with heave and pitch degrees of freedom. The dynamics system is modelled using Equation \eqref{MCK}.

\begin{equation}
    M\ddot{x}(t) + C\dot{x}(t) + K x(t)= \vec{F}_{tot}(t)
    \label{MCK}
\end{equation}
where $x$ is the 1D vector of displacements, $M$ the mass matrix, $C$ the damping matrix and $K$ the stiffness matrix. The vector $F_{tot}=(F_2,M_3)$ includes the total excitation forces and moments of the system.

The excitation forces and moments of the system are the integrated pressure  and viscous forces over the wall boundary:
\begin{equation}
    \begin{gathered}
        F_2 = \oint_{\partial B} pn_y + \left(\overline{\tau}\cdot\vec{n}\right)n_y \mathrm{dS} \\
        M_3 = \oint_{\partial B} \left(p\vec{n} + \overline{\tau} \cdot\vec{n}\right) \times
        \vec{r}\mathrm{dS} 
    \end{gathered}
    \label{excForces}
\end{equation}

\begin{figure}[H]
\centering
\includegraphics[width=0.7\textwidth]{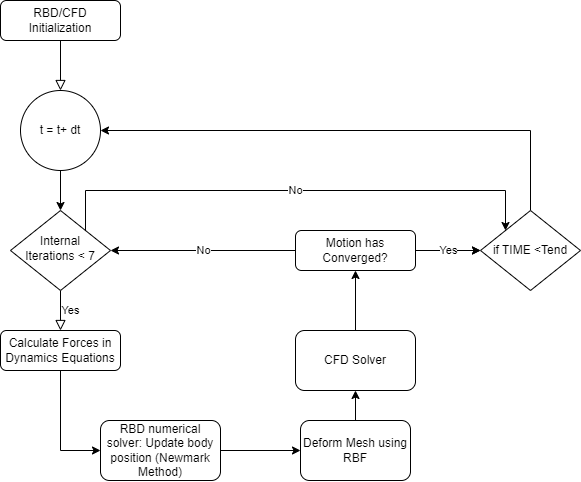}
	        \caption{Flow chart of the Fluid Solver-Rigid Body Dynamics (RBD) solver coupling. For each real time step, multiple internal iterations between the CFD and the Rigid Body Dynamics solver take place to ensure a strong coupling between the two.}
	        \label{fig:rbddeform}
\end{figure}

In FSI problems, the non-linear Equation \eqref{MCK} needs to be linearized to be solved iteratively. This linearization is performed during the pseudo-timesteps, and it allows the system to be solved more efficiently. In this work after each pseudo-timestep, the flow solver provides the forces and moments to the rigid dynamics solver (RBD), which  then numerically integrates (using the   the Newmark-$\beta$ method) the equation of motion  to compute the new position of the body. 
This iterative procedure results in a strong coupling between the fluid solver and the rigid body dynamics. A schematic representation of the algorithm used in FSI simulations is shown in  \autoref{fig:rbddeform}.

\section{Basic Numerical Setup \label{sec:numsetup}}

\begin{figure}
     \centering
     \includegraphics[width=0.7\textwidth]{ {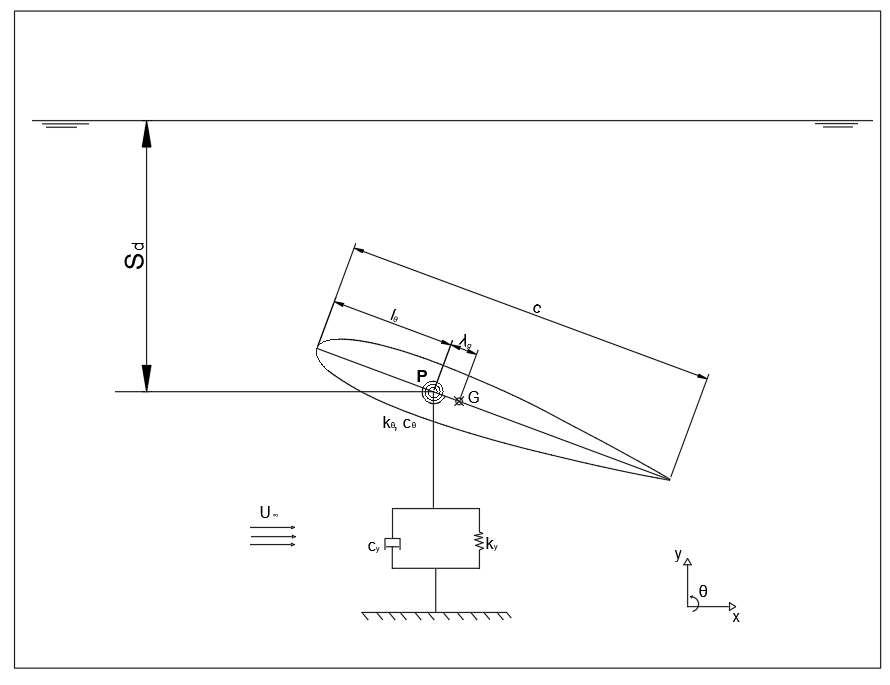}} 
     \caption{Schematic of the fully passive flapping foil device, operating in uniform current.}
     \label{fig:passive_foil}
  \end{figure}

The schematic of the fully passive flapping foil device is given in Figure \ref{fig:passive_foil}. In this work we examine the flapping foil in two dimensions and consequently three degrees of freedom (DOFs) are present, namely surge, heave and pitch. The device is comprised of a foil that is attached to linear springs and dampers in the heaving and pitching degrees of freedom, and its movement along the surge DOF is neglected. Fluid flow is horizontal, rightwards, as shown in the figure, while gravity acts vertically, downwards.  The corresponding stiffnesses for each spring are $k_y$ and $k_\theta$ and the damping coefficients are $c_y$ and $c_\theta$ for the heaving and pitching DOFs respectively. The pitching axis is located on the point $P$, as shown in the figure, which lies on the chord line of the foil, located at a distance $l_\theta$ from the leading edge. We also denote the distance $\lambda_G$ from the point $P$ to the center of gravity $G$ (CoG) of the device's moving parts. A positive value of $\lambda_G$ means that the center of gravity is located downstream of the pitching axis. The pitching axis generally differs from the CoG, which gives rise to the so called static imbalance defined in ~\eqref{imb}. This static imbalance couples the two DOFs and allows energy transfer between them. The parameter of submergence depth ($S_d$) is also defined in Figure~\ref{fig:passive_foil} as the distance from the foil's pitching axis to the free surface, when the heaving spring is not deformed.

\begin{align}
   \Lambda = \lambda_g  m_{\theta} 
   \label{imb}
\end{align}


Hydrodynamic and linkage loads (springs and dampers), along with inertial forces act on the foil. Applying Newton's second law, two nonlinear coupled differential equations arise, for the heaving and pitching DOFs, given in ~\eqref{rbd1} and ~\eqref{rbd2} \cite{Duarte2021}. 
	  
	 \begin{align} 
	   &m_y \ddot{y}+c_y \dot{y} + k_y y +\Lambda(\dot{\theta}^2sin\theta-\ddot{\theta}cos\theta)=F_y \label{rbd1} \\ 
	   &I_{\theta} \ddot{\theta} + c_{\theta} \dot{\theta} + k_{\theta} \theta -\Lambda(\ddot{y}  cos\theta) = M_{\theta}
            \label{rbd2}
	  \end{align}
  , where $m_y$ is the heaving mass and $I_{\theta} $ the moment of inertia of the foil with respect to the pitching axis, $(P,\overrightarrow{z})$. $m_y$ and $m_{\theta}$ might differ, since some mechanical components may participate only in one of the two motions. In the right side of the equations, $F_y$ is the hydrodynamic force acting on the heave direction and $M_\theta$ is the hydrodynamic moment with respect to the pitching axis. The non-dimensional form of the structural parameters used is given in Table~\ref{tb:nondim}.

  \begin{table}[H]
    \centering
    \caption{Definition of non-dimensional parameters used in the present study}
     \begin{tabular}{cccccc}
    	Parameter           & Definition   &  &  &   Parameter            & Definition  \\ \hline \hline
    	$Re$           & {\Large $\frac{U_{\infty}c}{\nu}$}               &  &  &  $m_\theta^*$           & \Large{$\frac{m_\theta}{\rho b c^2}   $} \vspace{2pt}   \\[2pt]  \hline 
        $m_y^*$  & \Large{$\frac{m_y}{\rho b c^2}$}  &  &  &  $\Lambda$ & $\lambda_g$  $m_{\theta}$ \vspace{2pt}              \\ \hline
        $k^*_{\theta}$ & \Large{$\frac{k_\theta}{\rho U_\infty^2 b c^2}$} &  &  &  $c^*_{\theta}$ & \Large{$\frac{c_\theta}{\rho U_\infty b c^3}$} \vspace{2pt} \\ \hline
        $k^*_{y}$      & \Large{$\frac{k_y}{\rho U_\infty^2 b}$}          &  &  &  $c^*_{y}$      & \Large{$\frac{c_y}{\rho U_\infty b c}$} \vspace{2pt}   \\ \hline
        $I^*_{\theta}$ & \Large{$\frac{I_{\theta}}{\rho b c^4}$}          &  &  &  $l^*_{\theta}$ & \Large{$\frac{l_{\theta}}{c}$} \vspace{2pt} \\ \hline
    \end{tabular} \label{tb:nondim}
\end{table}

\par The dampers effectively replicate the load on the foil from an electric generator, and since this energy can be calculated through $c_\theta$ and $c_y$ we proceed to define an efficiency coefficient to assess the foil performance. The hydraulic efficiency ($\eta$) is therefore defined as the integral of the ratio of power harvested divided by the hydraulic power available in the flow area S.
    
    \begin{align}
    	\eta=\frac{1}{\Delta t} \int_{t_0}^{t_0 +\Delta t} \frac{c_y  \dot{y}^2 + c_{\theta}\dot{\theta}^2 }{\frac{1}{2} \rho U_{\infty}^3 S  } dt
    	\label{eq:efficiency_passive}
    \end{align}
    
    , where $S$ is the maximum cross-sectional area swept by the foil, defined by the product of the foil's span, $b$, and the total vertical flow distance scanned by the foil.
    
    Another useful metric can be defined, normalizing the hydraulic power in terms of the projected surface of the foil ($b\times c$). Consequently the average power coefficient ($\overline{C}_p$) is defined as:
     \begin{align}
    	\overline{C}_p=\frac{1}{\Delta t} \int_{t_0}^{t_0 +\Delta t} \frac{c_y  \dot{y}^2 + c_{\theta}\dot{\theta}^2 }{\frac{1}{2} \rho U_{\infty}^3 b c  } dt
    	\label{eq:cp_passive}
     \end{align}
    

    
 Finally, the main parameters chosen for the fully passive flapping foil device, are given in their non-dimensional form in Table~\ref{tb:params} (taken from \cite{Duarte2019}).
\begin{table}[H]
    \centering
    \caption{Parameters used in the present study}
    \begin{tabular}{cccc}
    	Parameter           & Value      & Parameter            & Value    \\ \hline \hline
    	$Re$                & $6 \cdot 10^{4}$  &  $U_\infty$   & 1       \\ \hline 
    	$m*$                & 0.92              &  $\Lambda$           & 0.0065          \\ \hline
        $k^*_{\theta}$      & 0.071             &  $c^*_{\theta}$      & 0.052            \\ \hline
        $k^*_{y}$           & 0.72              &  $c^*_{y}$           & 0.93             \\ \hline
        $I^*_{\theta}$      & 0.0563            &  $l^*_{\theta}$      & 0.33             \\ \hline
    \end{tabular} \label{tb:params}
\end{table}

For the results presented the chord of the NACA0015 foil was set to $c=0.1$ and is considered to operate submerged in  water. Dimensionless force and power coefficients for each DOF are also defined in ~\eqref{dimf} in order to aid in the analysis of the results:

\begin{equation}\label{dimf}
\begin{aligned}
C_L = \frac{F_y(t)}{\frac{1}{2}\rho U_\infty^2bc}  \hspace{1cm} C_M = \frac{M_\theta(t)}{\frac{1}{2}\rho U_\infty^2bc^2} \\
C_{Py} = \frac{P_{F_y}}{\frac{1}{2}\rho U_\infty^3bc}  \hspace{1cm} C_{P\theta} = \frac{P_{M_\theta}}{\frac{1}{2}\rho U_\infty^3bc^2} 
\end{aligned}
\end{equation}

\begin{figure}[H]
			\centering
		\begin{subfigure}[b]{\textwidth}
			\includegraphics[width=\textwidth]{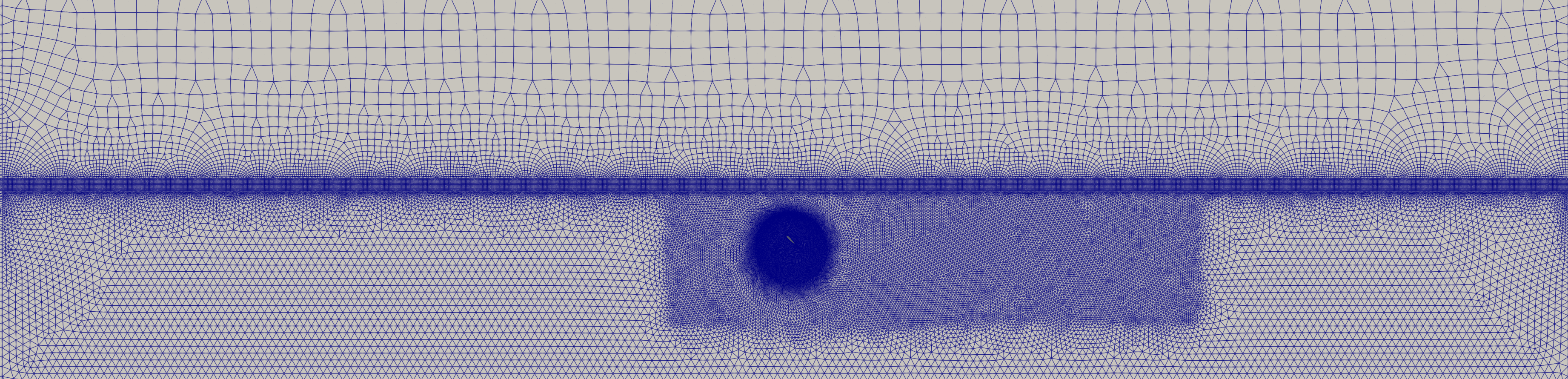}
		\end{subfigure}
	\caption{Mesh  overview  and  refinement zones for two-phase simulations \label{fig:freemesh}}
   	\end{figure}

Grid and time step convergence studies are provided in \cite{THEODORAKIS2022103674}. The grid and time step requirements are based on the comparison of the numerical results and measurements, without the effect of free surface.Since, in this work the foil operates beneath the free surface a new grid is generated  refined in the near free surface region. Nevertheless, the grid in the near airfoil region and the wake refinement zones remains approximately the same as the ones presented in the \cite{THEODORAKIS2022103674}. Regarding the free surface region discretization, it is horizontally refined to have at least 100 points per wavelength and 20 points per wave height ($H=0.2c$) (see \cite{ntouras2020coupled}). The  selected wavelength corresponds to generated waves with frequency close to the foil's motion frequency (see \autoref{sec:waves}). A mesh snapshot can be seen in \autoref{fig:freemesh}. At the right end of the computational domain a damping zone is employed. On the left end of the computational domain either a damping region or a wave generation zone is defined, depending on the desired simulation conditions.
	
The mesh used for the rest of the study consists of around $200000$ cells with refinements in the wake region to capture the vortex shedding. In the near wall region the first cell size is selected so that $y+ \le 1$ while the time step chosen is $\Delta t=0.0002s$. No initial perturbation is required to initiate the motion. Results are collected after enough periods have elapsed and the motion has sufficiently converged. The metrics used to evaluate performance are averaged across many periods.

\section{Results and Discussion}\label{sec:res}

In this section the aforementioned numerical setup of the fully passive flapping foil device is used to examine the influence of the free surface when it is either calm (\autoref{sec:depth}) or not (\autoref{sec:waves}). A parametric study is initially carried out, altering the submergence depth $S_d$. This analysis is repeated for various Froude (Fr) numbers, which is defined as:

\begin{equation}\label{eqn:frdef}
Fr = \frac{U_\infty}{\sqrt{g\cdot{c}}}
\end{equation}

The impact of these parameters to the performance is assessed, and their influence explained. Subsequently, the influence of monochromatic waves that propagate on the free surface is examined.

	\subsection{Effects of Submergence Depth for Various Froude Numbers}\label{sec:depth}

\begin{figure}[H]
    \centering
    \subfloat[\centering]{{\includegraphics[width=6.5 cm]{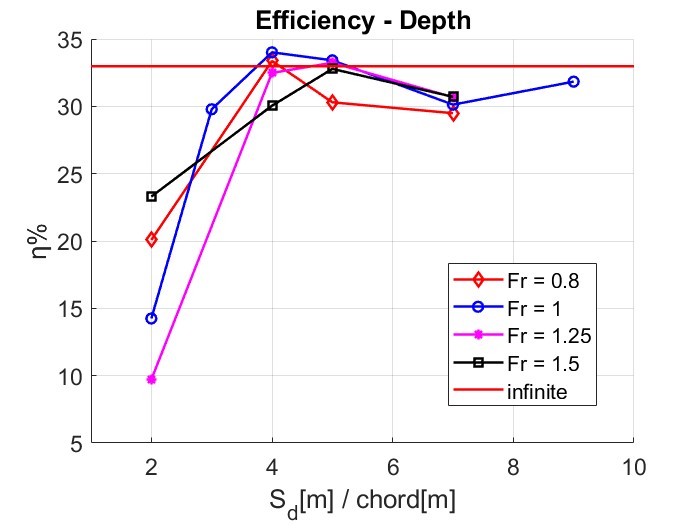} }}%
    \qquad
    \subfloat[\centering]{{\includegraphics[width=6.5 cm]{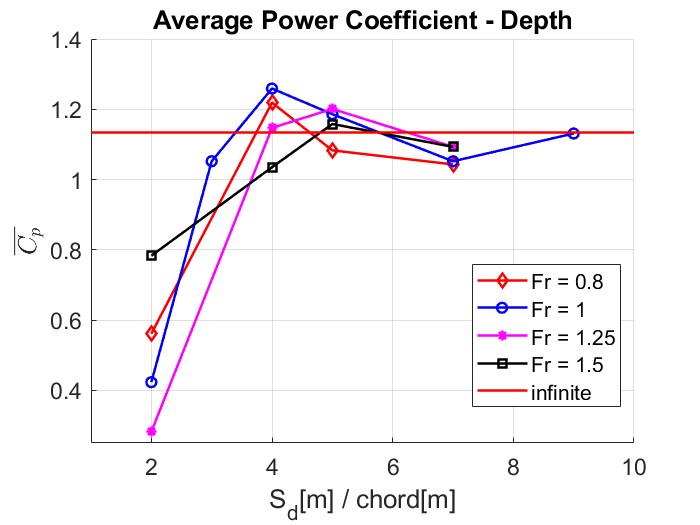} }}%
    \caption{a) Efficiency ($\eta$) and b) Average Power Coefficient ($\overline{C_p}$) curves for various Submergence Depths $S_d$. Each curve corresponds to a different Fr number. (compared to the infinite case: red horizontal line)}%
    \label{fig:fr}
\end{figure}

The effects of varying the submergence depth $S_d$ is examined in this section. This same analysis is carried out for multiple Fr numbers, Fr = 0.8, 1, 1.25, 1.5. The performance of the device is assessed using two metrics, efficiency $\eta$ and average power coefficient $\overline{C_p}$, described in \autoref{sec:num}. In Figure~\ref{fig:fr} (a) and (b) the results for all the cases examined are presented. Each curve in the $\eta-S_d$ and $\overline{C_p} - S_d$ diagrams, corresponds to a different Fr number. The horizontal red line represents the infinite depth case (no free surface influence) as a comparison to the other cases.  As a general trend, we notice that for low submergence depths, below 3 chord lengths (c), performance drops rapidly as the foil is very close to the free surface. For intermediate $S_d$, a maximum appears for all the Fr numbers, where $\eta$ and $\overline{C_p}$ are as high or even higher than the infinite depth case. Increasing $S_d$ further, performance drops again until it  approaches asymptotically the infinite depth case.

Results indicate that there exists an optimal submergence depth where the calm free surface not only does not decrease performance relative to the infinite case, but also a slight increase is evident. The exact depth $S_d$ that the maximum occurs depends on the Fr number. As \autoref{fig:fr} indicates even though the Fr number can affect performance the efficiency maximum can be found at intermediate $S_ds$ (between $4c-5c$). To further study the effect of the submergence depth we  isolate the investigation at an single Fr number.

\subsubsection{Effects of Submergence Depth for Fr = 1}

\begin{figure}[H]
    \centering
    \subfloat[\centering]{{\includegraphics[width=6.5cm]{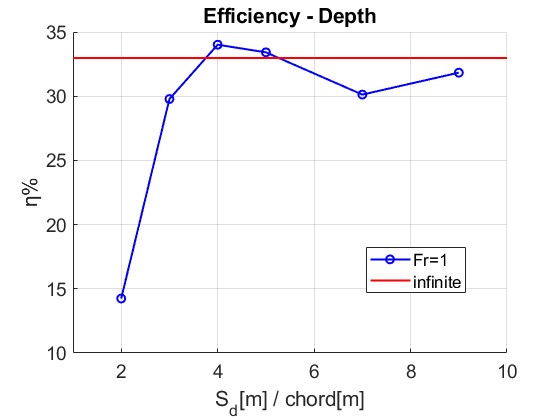} }}%
    \qquad
    \subfloat[\centering]{{\includegraphics[width=6.5cm]{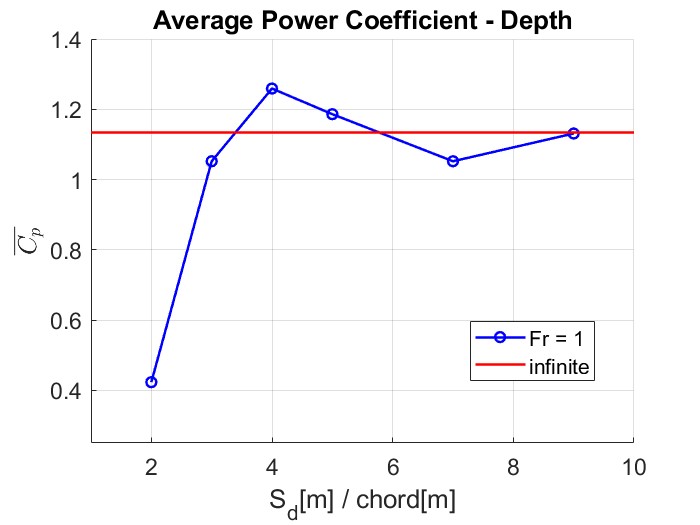} }}%
    \caption{Fr = 1: a) Efficiency $\eta$ and b)Average Power Coefficient $\overline{C_p}$ for various submergence depths}%
    \label{fig:fr1}
\end{figure}

To better understand the effect of submergence depth and the underlying fluid mechanics phenomena we focus our study to   Fr = 1. In Figure~\ref{fig:fr1} (a) the $\eta-S_d$ and in Figure~\ref{fig:fr1} (b) $\overline{C_p}-S_d$ plots for Fr = 1 are isolated. The submergence depths examined for the specific Fr number are $S_d=2,3,4,5,7$ and $9$ chord lengths.

The main metrics for each submergence depth, for Fr = 1 are presented in Table~\ref{tab:fr1}, where h* and $\theta$* are heaving and pitching amplitudes (heave is non-dimensional by chord length c and $\theta$* is in degrees) and T* is non-dimensional period.

\begin{table}[H]
\begin{center}
\centering
\begin{tabular}{||c c c c c c c c||} 
 metrics/$S_d$ & 2c & 3c & 4c & 5c & 7c & 9c & Infinite \\ [0.5ex] 
 \hline\hline
 $h*$ & 1.26 & 1.45 & 1.44 & 1.40 & 1.40 & 1.41 & 1.32 \\ 
 \hline
 $\theta$ & 47.7 & 66.3 & 71.8 & 70.7 & 65.7 & 67.9 & 72.2 \\
 \hline
 $\eta\%$ & 14.25 & 29.67 & 34.10 & 33.42 & 30.13 & 31.83 & 32.98 \\
 \hline
 T* & 12.6 & 9.0 & 8.3 & 8.4 & 8.9 & 8.6 & 8.1  \\
 \hline
 $\overline{C_p}$ & 0.42 & 1.06 & 1.26 & 1.19 & 1.05 & 1.13 & 1.13 \\ [1ex] 
 \hline
\end{tabular}
\caption{Fr = 1: Main Metrics for various Submergence Depths.}
\label{tab:fr1}
\end{center}
\end{table}

As \autoref{tab:fr1} suggests , the influence of the free surface is not straightforward. Efficiency and Power Coefficient seem to have a similar behaviour. For high $S_d$, such as $S_d=7c$ and $9c$ the presence of the free surface slightly reduces the performance of the device. The influence is small as the infinite case is approached for high $S_d$. For very low $S_ds$ close to $2c$ performance reduces rapidly. The hydrofoil is very close to the free surface and large amounts of energy are expended to the formation of waves. Other reasons, relating to the behaviour of the vortices around the foil affect its performance, and are discussed later in this section. 

\begin{figure}[H]
    \centering
    \subfloat[\centering]{{\includegraphics[width=6.5cm]{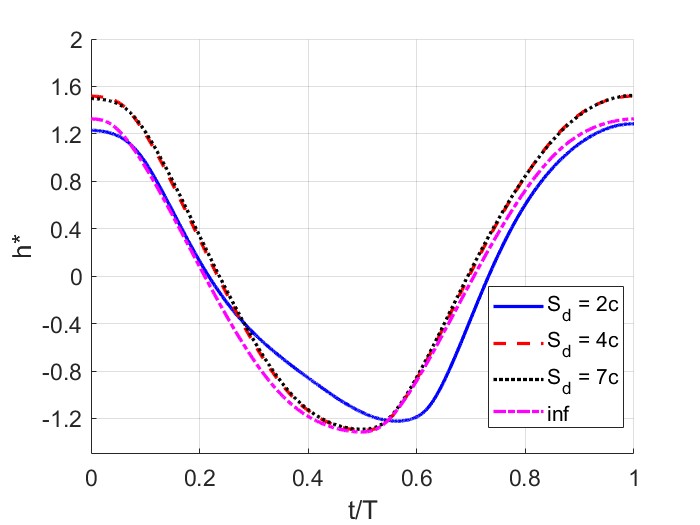} }}%
    \qquad
    \subfloat[\centering]{{\includegraphics[width=6.5cm]{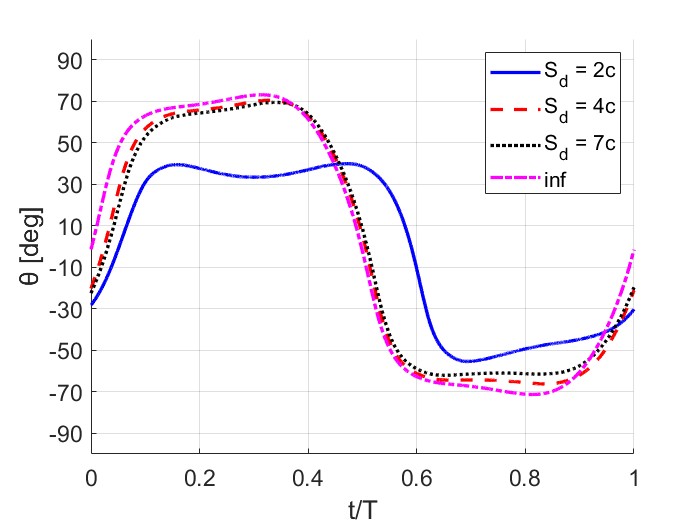} }}%
    \caption{a) h* and b) $\theta$* for a single period for different $S_d$s}%
    \label{fig:kin}
\end{figure}

In \autoref{fig:kin} (a) and (b) the heaving and pitching motions can be found, for submergence depths $S_d=2c, 4c, 7c$. It is evident that  the heaving amplitude ($h^*$) is not significantly affected when the foil is at low $S_ds$. The pitching amplitude, in contrast, is significantly decreased. 

\begin{figure}[H] 
	\centering
	\includegraphics[width = 0.5\textwidth]{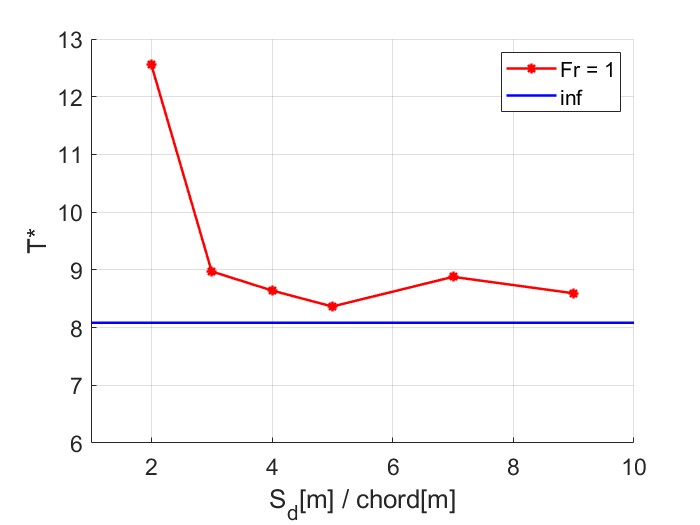}
	\caption{T* for different $S_d$ relative to the infinite case.}\label{fig:Per}
\end{figure}

Additionally, in Figure~\ref{fig:Per} the non-dimensional period $T^*$-$S_d$ is presented and compared to the infinite case. It is clear, that the period (T*) is greatly increased when the foil operates near the free surface ($S_d=2c, 3c$) This increased period, means less energy is harvested from the foil, thus the drop in performance.  For intermediate $S_ds$, such as $4c$ and $5c$, a peak forms, and T* is very close to the reference infinite case.

\textbf{Depletion of energy in waves}

Free surface directly affects the performance of the flapping foil as energy is transferred from the foil to the creation of gravity waves. As the foil approaches the surface, larger  waves are formed, which means that more energy is lost in wave-making. In essence, the pressure differences that the foil creates lend part of their energy to the creation of waves leading to the reduction of lift. Additionally, the formed wave system can affect the pressure distribution on the foil and thus the resulting motion.

\begin{figure}[H] 
	\centering
	\includegraphics[width = 0.8\textwidth]{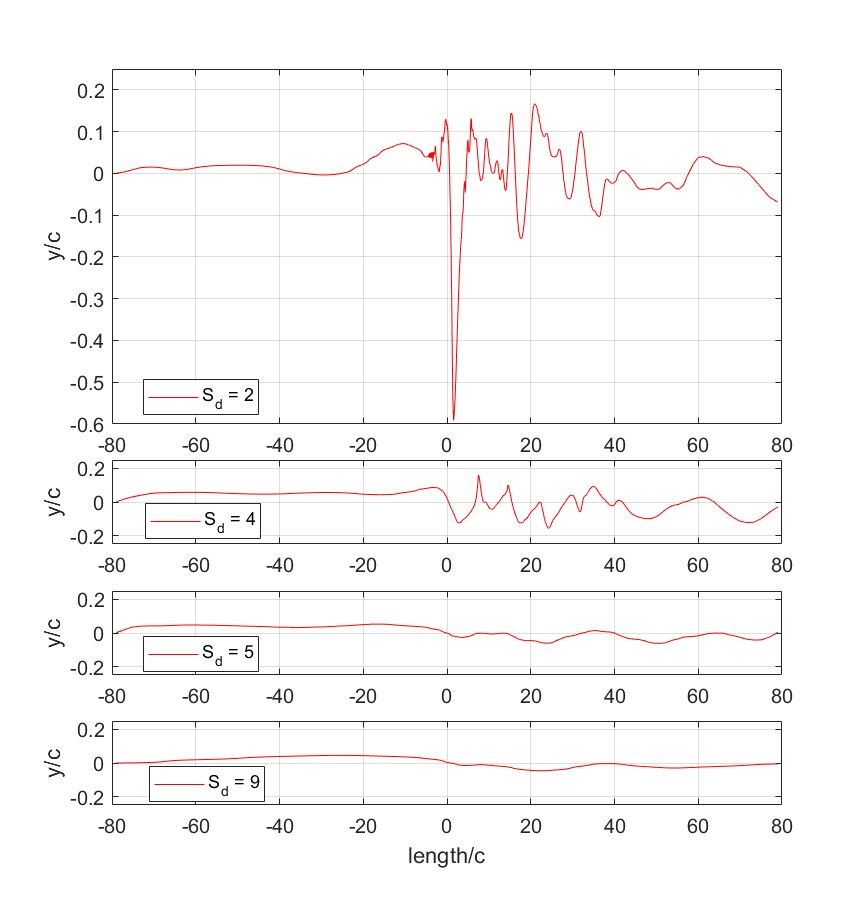}
	\caption{Free surface disturbance for the whole computational domain for various submergence depths (foil L.E. is at 0)}\label{fig:dist}
\end{figure}

\begin{figure}[H] 
	\centering
	\includegraphics[width = 1\textwidth]{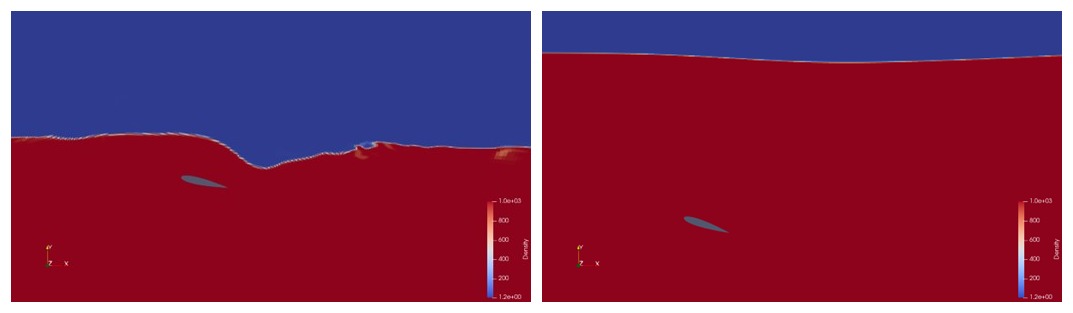}
	\caption{Free surface visualisation near the foil region for $S_d$ = 2c (left) and $S_d$ = 4c (right)(red: water phase and blue: air phase).}\label{fig:dens}
\end{figure}

The free surface  disturbance for the various $S_ds$ can be seen in \autoref{fig:dist}. We chose to present the free surface when the foil is at its uppermost position. As expected, the disturbances are larger for low submergence depths, reaching wave amplitudes up to $0.6c$.  The free  surface is also visualized by the density plot in \autoref{fig:dens}, comparing $S_d=2c$ with $S_d=4c$ when the foil is at its uppermost position, where the disturbance in the low $S_d$ case is particularly high. 

It is notable,however,  that even though at $S_d = 4c$ significant waves are formed in the free surface the  performance of the foil is slightly increased. This signifies that free surface wave making is not the only factor affecting performance.

\textbf{Free Surface Flow Constriction}

 The second significant effect that affects performance when the foil operates near the free surface is   caused due the constriction of the flow. The pressure drop due to the constriction of the flow, results in increased velocities in the region. This increased velocity  along with the influence of the disturbance of the free surface, affect the pressure on the upper side of the foil.


 


\begin{figure}[H] 
	\centering
	\includegraphics[width = 0.5\textwidth]{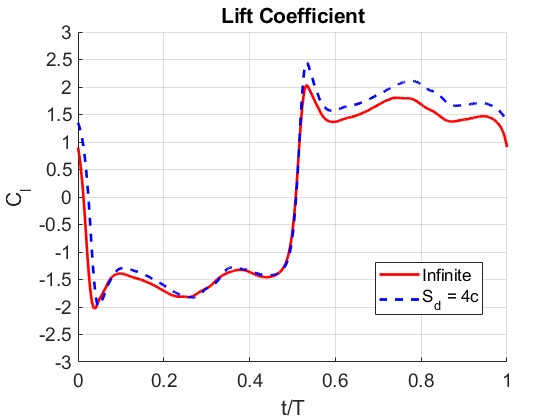}
	\caption{Lift Coefficient ($C_L$) comparison for a single period.}\label{fig:CL}
\end{figure}

At a certain depth, constriction of the flow can be beneficial for energy extraction. As the foil heaves upwards, pressure is decreased further on its upper side compared to the infinite case, causing the Lift force to increase. This means that more energy is harvested from the device. As the foil moves downwards, this effect is reversed, which means that lift is decreased compared to the infinite depth case. This is illustrated in Figure~\ref{fig:CL} for a single period, where the infinite depth case Lift Coefficient ($C_l$ is contrasted to that of the $S_d = 4c$ case. In the first half-period, the magnitude of lift is slightly less but in the second half it is greater. It can be seen that the increase in the absolute value of the $C_l$ on the second half of the period (upstroke) is greater than the decrease on the first half (downstroke). This explains why there is a net positive effect on the foil’s performance, at the particular $S_ds$. 

\begin{figure}[H]
    \centering
    \subfloat[\centering]{{\includegraphics[width=6.5cm]{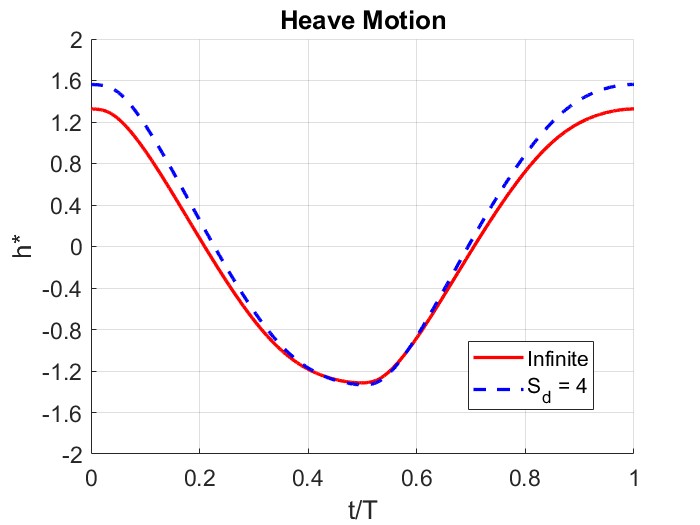} }}%
    \qquad
    \subfloat[\centering]{{\includegraphics[width=6.5cm]{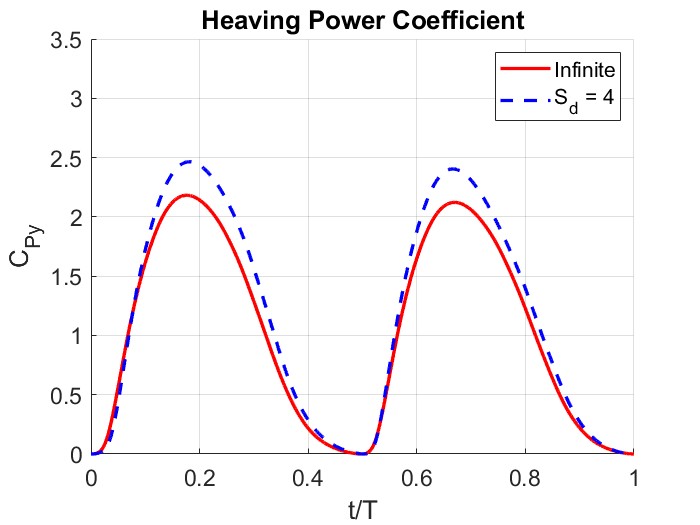} }}%
	\caption{a)h*(t*) and b) $C_{Py}$ comparison for $S_d= 4c$ and infinite case.}
    \label{fig:brn}
\end{figure}

The increased $C_l$ during the upstroke, increases the amplitude of the heaving motion, as depicted in Figure~\ref{fig:brn} (a) (while it is the same at the downstroke). Figure~\ref{fig:brn} (b) depicts the heaving power coefficient, where it is evident that power extraction is increased for $S_d = 4c$, throughout the period. Energy extraction from the pitching motion is much less than that of the heaving motion, and it does not differ significantly between these cases. 

\textbf{Motion Synchronisation}

Synchronisation of the heaving and pitching motion is very important in the search of structural parameters in order to achieve good performance. This device is not active, meaning that the motion is not prescribed, so the motion cannot be explicitly set. In essence, the pitching motion is responsible for the suitable positioning of the foil, so that the maximum amount of power can be harvested by the heaving motion. Consequently, when the motion in the two DOFs are not well synchronised, harvested energy decreases. 

\begin{figure}[H] 
	\centering
	\includegraphics[width = 0.7\textwidth]{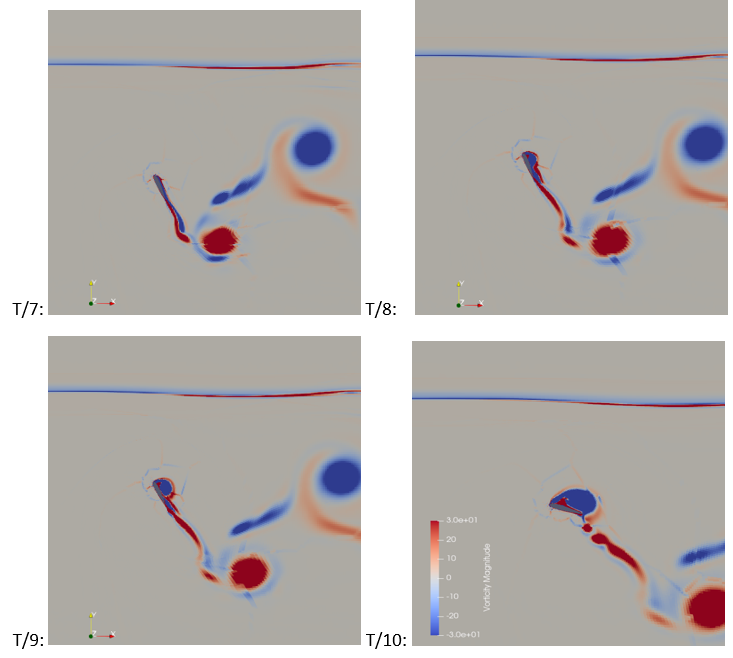}
	\caption{Vorticity contours for successive positions of the foil during the upstroke.}\label{fig:vort}
\end{figure}

As described by \cite{birch2004force}, a large Leading Edge Vortex (LEV) is formed close to the leading edge of the foil, as the foil goes through dynamic stall, which is detached when certain conditions are met. The vorticity of the flow during an upstroke for the $S_d = 4$ case is shown in Figure~\ref{fig:vort}. It can be seen, that the  LEV forms, detaches, and afterwards is convected downstream along the upper surface of the body. Apart from the structural parameters, the kinematics of the LEV is a significant factor that drives the passive foil motion.

\begin{figure}[H] 
	\centering
	\includegraphics[width = 1\textwidth]{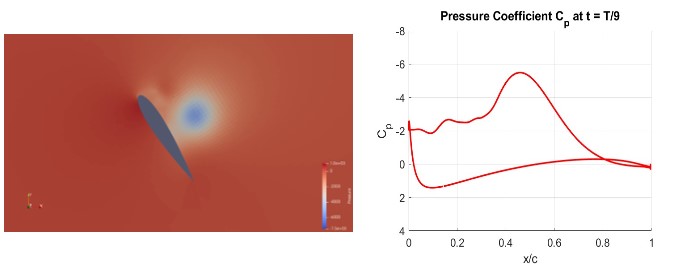}
	\caption{Pressure contour at T/9 (a) and the corresponding pressure coefficient Cp plot (b)}\label{fig:LEV}
\end{figure}

 Figure~\ref{fig:LEV} (a) shows the pressure contour at $T/9$ of the motion cycle. The LEV can be identified as the  low pressure area. It's effect on the foil loading can be seen in  Figure~\ref{fig:LEV} (b) that shows the corresponding pressure coefficient (Cp) plot at T/9 the same time instant. It is clear that the location of the LEV results in a lower pressure region acting on the foil.
 
 The LEV induced pressure drop results in a large counter-clockwise moment that  acts on the hydrofoil. Consequently, the hydrofoil changes orientation so that the downstroke will begin. This mechanism determines the pitching motion and affects synchronisation.

When the hydrofoil is located close to the free surface, a delay in the pitching motion is noticed which is caused by a delay of the shedding of the LEV. The main cause of this delayed shedding is best described by \cite{sheridan1997flow}. Simulations of cylinders close to the free surface showed a clear delay in the shedding frequency, caused by the restriction of the fluid flow by the free surface, which restricts the supply of fluid in the region of the vortex.   

\begin{figure}[H] 
	\centering
	\includegraphics[width = 0.6\textwidth]{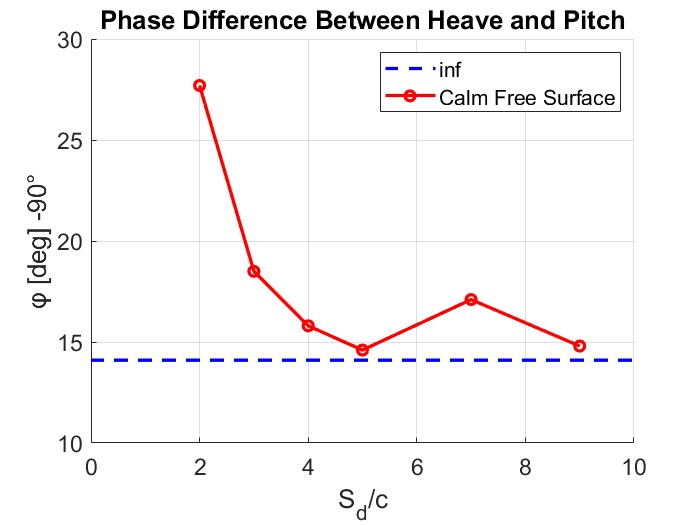}
	\caption{Heave and Pitch Phase difference (-90°) for various depths (Fr = 1)}\label{fig:phs}
\end{figure}

To assess the synchronisation of the heaving and pitching motions, as $S_d$ varies, Fourier analysis was conducted to find their phase difference. These were compared then to the infinite case. The heaving and pitching motion time series were analysed and their corresponding phases were subtracted in order to find their phase difference. Results are presented in Figure~\ref{fig:phs}.

These results inversely correlate with the results for the performance ($\eta$ and $\overline{C_p}$) of the foil in Figure~\ref{fig:fr1} (a) and (b). For $S_ds$ where performance is maximised  ($S_d= 4c, 5c$), phase difference is close to the infinite case. For $S_d=7c$ phase difference is larger, and performance drops. For lower submergence depths ($S_d=2c,3c$) phase difference becomes much larger as performance drops rapidly. This shows that when the foil is located very close to the free surface, the synchronisation is negatively affected, due to the delay in the shedding of the LEV. In essence, due to the bad synchronisation, the loads applied on the hydrofoil by the flow are lower, and the harvested energy decreases.  

\textbf{Heaving Motion Resonance}

The heaving motion is responsible for the bulk of the energy harvested from the incoming flow. An increase in the amplitude of the heaving motion corresponds to an increase in the harvested energy. As this is an oscillating system, synchronisation between the loads applied and the natural frequency of the system would result in increased amplitude. A Fourier analysis was done for the heaving velocity time-series $\dot{y}(t)$, as velocity is directly related to the amount of harvested power and thus the coefficients $\eta$ and $\overline{C_p}$. The predominant frequency is then compared to the natural frequency of the heave motion. The undamped system's natural frequency is given by equation~\ref{eqn:eq1}.

\begin{equation}\label{eqn:eq1}
\omega_n = \sqrt{\frac{k_y}{m_y}}
\end{equation}

However, as this is a damped system, the natural frequency is reduced, so the system will oscillate freely with a new damped natural frequency which depends on the damping ratio $\zeta$ (equation~\ref{eqn:eq2}):

\begin{equation}\label{eqn:eq2}
\omega_d = \omega_n\sqrt{1-\zeta^2}
\end{equation}

, where: \[\zeta = \frac{c_y}{2\sqrt{mk_y}}\]

\begin{figure}[H] 
	\centering
	\includegraphics[width = 1\textwidth]{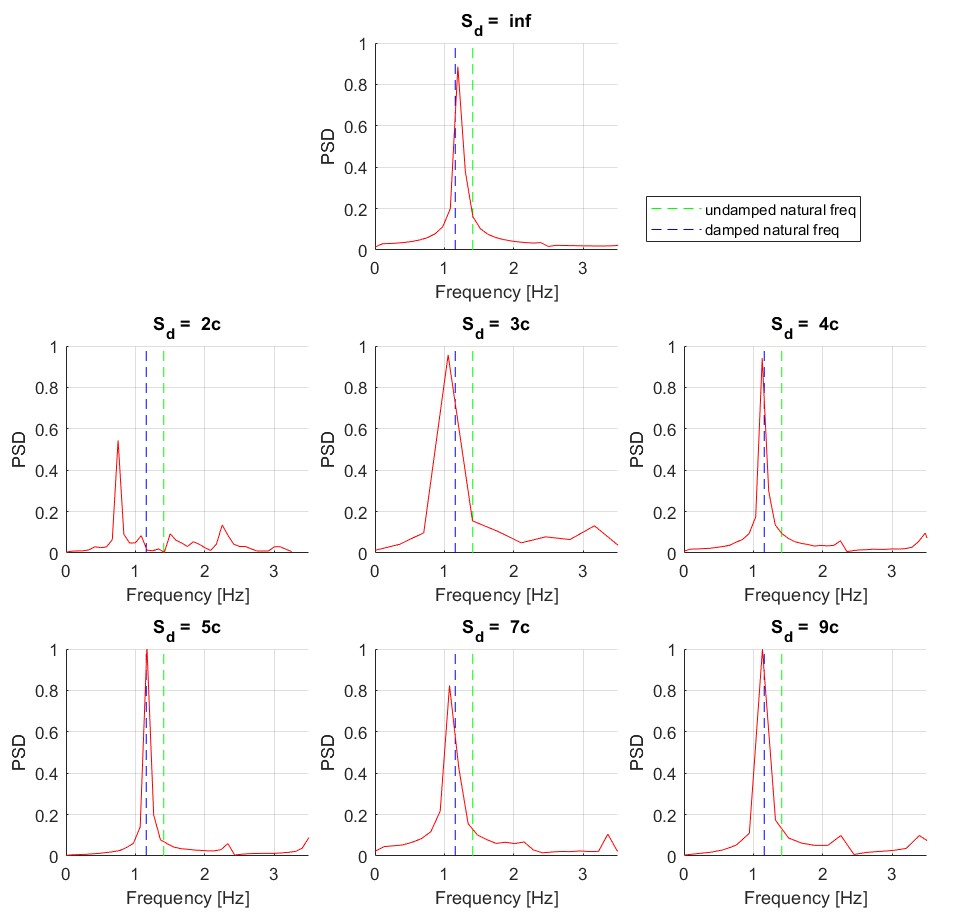}
	\caption{Power Spectral Density-f(Hz) diagrams from Fourier analysis of the $\dot{y}(t)$ timeseries for different depths. Vertical lines show the undamped and damped natural frequencies.}\label{fig:psd}
\end{figure}

The frequency – Power Spectral Density (PSD) diagrams of the Fourier Analysis are presented in Figure~\ref{fig:psd}.

The results of the Fourier analysis indicate that good performance is correlated with the resonant frequency of the heaving motion and its proximity to the damped natural frequency. For submergence depths $S_d= 4c, 5c, 9c$, where performance is high, the corresponding peaks in the diagrams almost match the natural frequency of the damped system $f_d$. In contrast, for $S_d$ = 7c there is a slight difference, as performance is also slightly lower. As for the case $S_d$ = 3c and especially for $S_d$ = 2c, the frequency drops by a very significant amount due to the reduced loads close to the free surface, so that the absence of resonance causes the large drop in performance. Although the device was well-tuned in infinite fluid, so that resonance occurs, this is negatively affected close to the free surface as the oscillation frequency drops rapidly.

\subsection{Effects of monochromatic Waves}{\label{sec:waves}}

After examining the effects of a calm free surface on the performance of the fully passive flapping foil device, a parametric study is undertaken to examine the influence of monochromatic waves. The device is expected to operate near the free surface and consequently waves can affect its performance.


In particular, a parametric study takes place varying the frequency of these waves.The waves have frequency ($\omega_w$),wavelength ($\lambda$) and wavenumber ($k$). The waves are generated using streamfunction theory \cite{Fenton1988}.They are generate from the left and of the computational domain using a wave-generation zone and propagate towards the passive foil.



Following \autoref{sec:depth} the submergence depth $S_d=4c$ is selected, since, at this depthefficiency is maximised. Reynolds number is still $6\cdot10^4$ and Fr = 1 was chosen. The device's structural parameters and the mesh are also kept the same as in \autoref{sec:depth}. The following ratios (\autoref{tab:omg}) of $\omega_w/\omega_f$ are investigated.


\begin{table}[H]
\begin{center}
\centering
\begin{tabular}{c|c|c } 
 $\omega_w /\ \omega_f$ & $\lambda$ & Heave Amplitude h* \\ 
 \hline
 calm & \textbf{--} &  1.44 \\
 0.5 & $77.3c$&  1.46 \\
 0.8 & $38.9c$&  1.50 \\
 0.9 & $32.4c$&  1.44 \\
 1.0 & $26.1c$&  1.48 \\
 1.1 & $23.9c$&  1.48 \\
 1.2 & $21.0c$&  1.48 \\
\end{tabular}
\caption{The investigated wave frequencies and wavelengths (in airfoil chords), and the predicted non-dimensional heaving amplitudes h* = h/c.}
\label{tab:omg}
\end{center}
\end{table}

The wave height $H = 2A$ (A: amplitude) has to be comparable to the disturbance of the free surface caused by the foil. For this reason, it was chosen equal to $0.2c (A = 0.1c)$ for all cases. The tank has a depth of 18c which is large relative to the wavelengths . This means that wave propagation happens in deep water conditions. The corresponding wavelengths can be seen in \autoref{tab:omg}. Finally, it is noted that the selected wave height classifies  the waves in the non-linear region.

\begin{figure}[H]
    \centering
    \subfloat[\centering]{{\includegraphics[width=6.5cm]{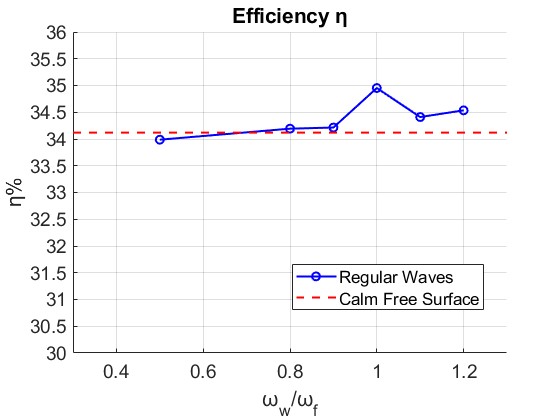} }}%
    \qquad
    \subfloat[\centering]{{\includegraphics[width=6.5cm]{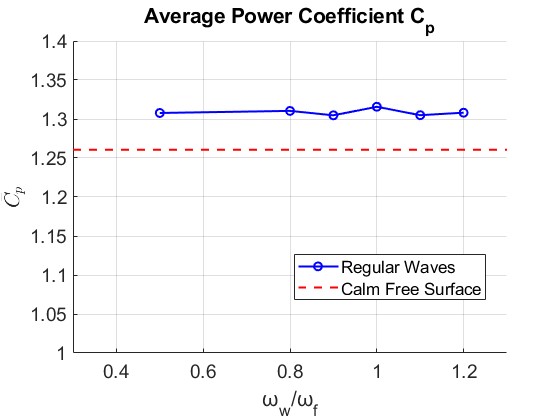} }}%
	\caption{a) $\eta$ and b) $\overline{C_p}$ – Encounter Frequency/Wave frequency $\omega_w/\omega_f$}
    \label{fig:wv}
\end{figure}

In \autoref{fig:wv} the performance of the passive foil operating in waves can be found. Figure~\ref{fig:wv} (a), indicates that the presence of waves does not affect efficiency significantly, except for the case of $\omega_w/\omega_f = 1$, where a slight increase of about 0.8$\%$ is witnessed. This indicates that resonance of the two motions increases power extraction.

On the contrary, in Figure \ref{fig:wv} (b) it is seen that the average power coefficient is increased for all cases, by about the same amount, 4.5$\%$. This suggests that the harvested power due to the presence of waves is increased. This is due to  to the increased heaving amplitude when operating in waves as indicated  in \autoref{tab:omg}. This also explains the smaller changes in the efficiency of the device since $\eta$ is normalized by the swept area.




\section{Conclusions}\label{sec:conclusions}
In conclusion, the performance of a fully passive flapping foil device for energy harvesting was investigated in a free surface flow through a series of numerical simulations. We used a strongly coupled FSI algorithm to examine the effects of varying submergence depths and the impact of monochromatic waves on the foil's performance.

The results showed that the fully passive flapping foil device can achieve high efficiency for submergence depths between $4c$ and $9c$. It is notable that there is an "optimum" submergence depth where the flapping foil performance is maximised. Simulations for different Froude numbers showed the same trends, with the optimum always found between submergence depths of $4c$-$5c$. The performance was found to be correlated with the resonant frequency of the heaving motion and its proximity to the damped natural frequency. The influence of the large leading edge vortex was also found to be important.

Finally, we investigated the effects of regular waves on the foil's performance. We found that waves with a frequency close to that of the natural frequency of the flapping foil aided energy harvesting, increasing the energy extraction. As the passive foil operates in waves, its swept area increases, resulting in an increase in the power extracted from the waves. However, the foil's efficiency remains almost constant. 

Overall, this study contributes to a better understanding of the performance of fully passive flapping foil devices for energy harvesting in free surface flows and provides insights that could be useful for future design improvements. The next steps involve examining the performance of the fully passive flapping foil device in more complex and realistic ocean conditions, which include considering the foil operating in a wave spectrum and a 3D configuration. 


\section*{Acknowledgments}
This work was supported by computational time granted from the Greek Research \& Technology Network (GRNET) in the National HPC facility - ARIS - under project  "DYNASEA" with ID pr012019. The computational grids were generated using the ANSA pre-processor of BETA-CAE Systems.

\bibliographystyle{elsarticle-num} 
\bibliography{main.bib}
  




\end{document}